\documentclass{article}
\usepackage{cite}
\usepackage{amsmath,amssymb,amsfonts}
\usepackage{algorithmic}
\usepackage{graphicx}
\usepackage{textcomp}
\usepackage{xcolor}
\usepackage{textgreek}
\usepackage{mathtools}
\usepackage{engord}
\usepackage{url}
\usepackage{enumitem}
\usepackage{calc}
\usepackage{subfig}

\def\BibTeX{{\rm B\kern-.05em{\sc i\kern-.025em b}\kern-.08em
    T\kern-.1667em\lower.7ex\hbox{E}\kern-.125emX}}

\providecommand{\keywords}[1]
{
  \small	
  \textbf{\textit{Index Terms---}} #1
}

\begin{document}

\title{Random Number Generator Attack against the Kirchhoff-Law-Johnson-Noise Secure Key Exchange Protocol\\
}

\author{Christiana Chamon\textsuperscript{1}, Shahriar Ferdous\textsuperscript{2}, and Laszlo Kish\textsuperscript{3}\\
\textit{Department of Electrical and Computer Engineering} \\
\textit{Texas A\&M University}\\
College Station, TX 77843-3128, USA \\
cschamon@tamu.edu, ferdous.shahriar@tamu.edu, laszlokish@tamu.edu\\
ORCiD: \textsuperscript{1}0000-0003-3366-8894, \textsuperscript{2}0000-0001-5960-822X, \textsuperscript{3}0000-0002-8917-954X}

%get rid of the date
\date{}

\maketitle

\begin{abstract}
This paper introduces and demonstrates two new attacks against the Kirchhoff-Law-Johnson-Noise (KLJN) secure key exchange scheme. The attacks are based on random number generators with compromised security. First we explore the situation in which Eve knows the seed of both Alice's and Bob's random number generators. We show that in this situation Eve can crack the secure key bit within a fraction of the bit exchange period even if her current and voltage measurements have only a single bit of resolution. In the second attack, we explore the situation in which Eve knows the seed of only Alice's random number generator. We show that in this situation Eve can still crack the secure key bit but she needs to use the whole bit exchange period for the attack. The security of the KLJN key exchange scheme, similarly to other protocols, necessitates that the random number generator outputs are truly random for Eve.
\end{abstract}

\keywords{random number generator, secure key exchange, unconditional security}

\section{Introduction}
\label{intro}
\subsection{On Secure Communications}
\begin{figure}[htbp]
\centerline{\includegraphics[width=8.25cm]{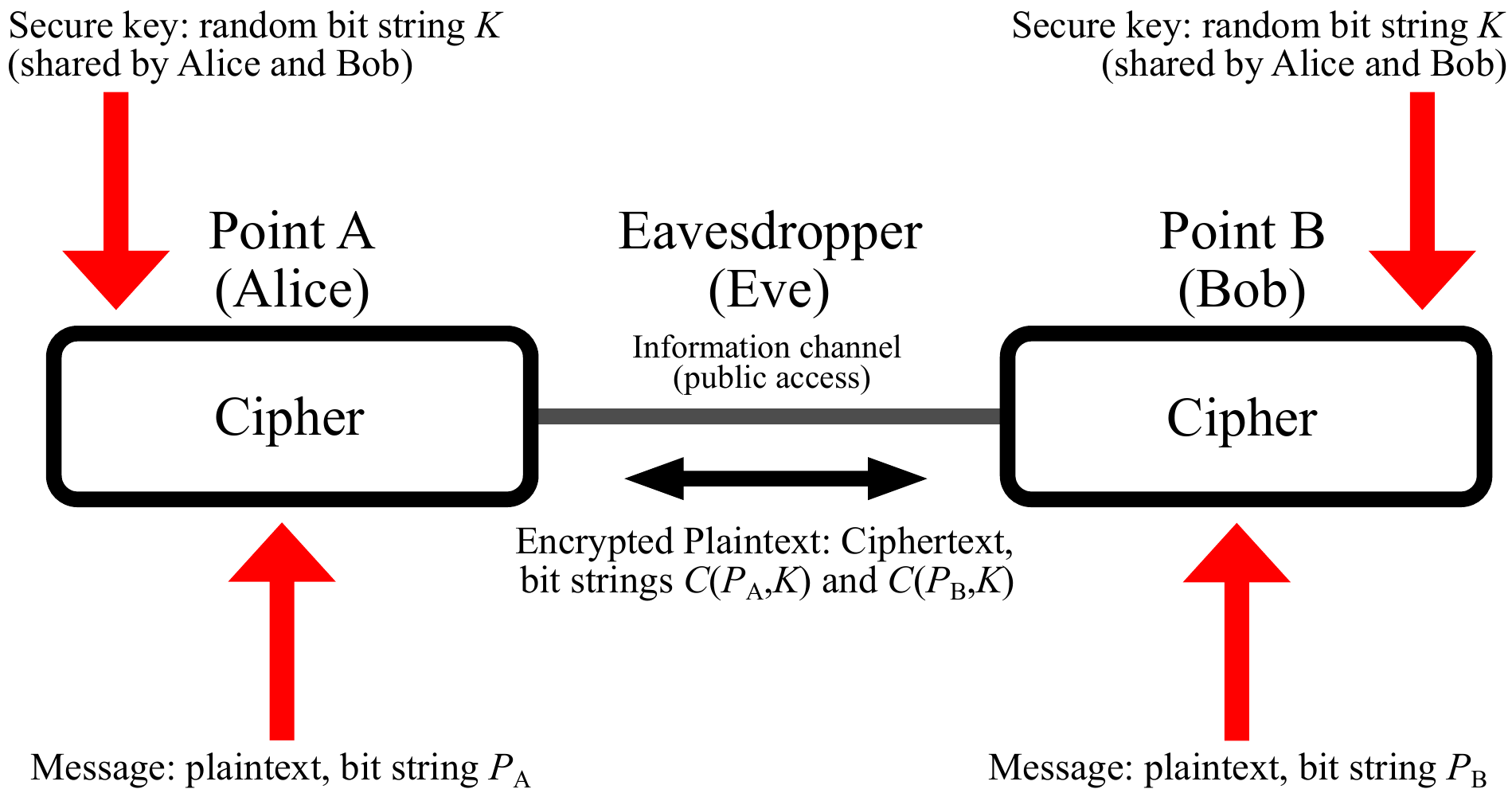}}
\caption{Symmetric-key cryptography \cite{b1}. Alice and Bob use ciphers to exchange secure keys, or a string of bits, through a public channel. The ciphers encrypt plaintext, or convert it into ciphertext. The secure key is represented by $K$, the plaintext messages of Alice and Bob are represented by $P_\mathrm{A}$ and $P_\mathrm{B}$, respectively, and the ciphertext is represented as a function of $P$ and $K$, or $C$($P$,$K$).}
\label{symmetric}
\end{figure}

%What is the topic of the paper? Why should I as a reader care about that topic? What has been done previously? What is the novelty that this paper brings to the table?
One way to establish the security of a communication is through encryption, that is, the conversion of plaintext into ciphertext via a cipher \cite{b1}. Fig.~\ref{symmetric} provides the general scope of symmetric-key cryptography \cite{b1}. The key is a string of random bits and both communicating parties Alice and Bob use the same key and ciphers to encrypt and decrypt their plaintext.

For a plaintext message $P$ and a secure key $K$, the encrypted message, or the ciphertext $C$, is a function of $P$ and $K$, that is,

\begin{equation}
C=C(P,K).\label{ciphertext}
\end{equation}

In symmetric-key cryptography, for decryption, the inverse operation is used:

\begin{equation}
P=C^{-1}[C(P,K),K].\label{plaintext}
\end{equation}

\noindent Because the secure keys must be the same at the two sides (shared secret), another type of secure data exchange is needed before the encryption can begin: the secure key exchange, which is the generation and distribution of the secure key over the communication channel. Usually, this is the most demanding process in the secure communication because the communication channel is accessible by Eve thus the secure key exchange is itself a secure communication where the cipher scheme shown in Fig.~\ref{symmetric} cannot be used. Eve records the whole communication during the key exchange, too. She knows every detail of the devices, protocols, and algorithms in the permanent communication system (as stated by Kerchhoff’s principle \cite{b2}/Shannon's maxim), except for the key. In the ideal case of perfect security, the key is securely generated/shared, immediately used by a One Time Pad \cite{b3}, and destroyed after the usage. In practical cases, usually there are deviations from these strict conditions, yet the general rule holds: A secure system cannot be more secure than its key.

The key is assumed to be generated from truly random numbers. Any predictability of the key leads to compromised security \cite{b3}. In this paper, we demonstrate attacks on the unconditionally secure Kirchhoff-Law-Johnson-Noise (KLJN) secure key exchange based on compromised random number generators (RNGs).

\subsection{The KLJN Scheme}
\label{KLJN}
The KLJN scheme \cite{b41,b42,b43,b44,b45,b46,b47,b48,b49,b50,b51,b52,b53,b54,b55,b56,b57,b58,b59,b60,b61,b62,b63,b64,b65,b66,b67,b68,b69,b70,b71,b72,b73,b74,b75,b76,b77,b78,b79,b80,b81,b82,b83,b84,b85,b86,b87,b88,b89,b90,b91,b92} is a statistical physical scheme based on the thermal noise of resistors. It is a classical (statistical) physical alternative of Quantum Key Distribution (QKD). Note, in papers  \cite{b3,b4,b5,b6,b7,b8,b9,b10,b11,b12,b13,b14,b15,b16,b17,b18,b19,b20,b21,b22,b23,b24,b25,b26,b27,b28,b29,b30,b31,b32,b33,b34,b35,b36,b37,b38,b39,b40}, important criticisms and attacks are presented about QKD indicating some of the most important difficulties of unconditionally secure quantum hardware and their theory.

\begin{figure}[htbp]
\centerline{\includegraphics[width=8.25cm]{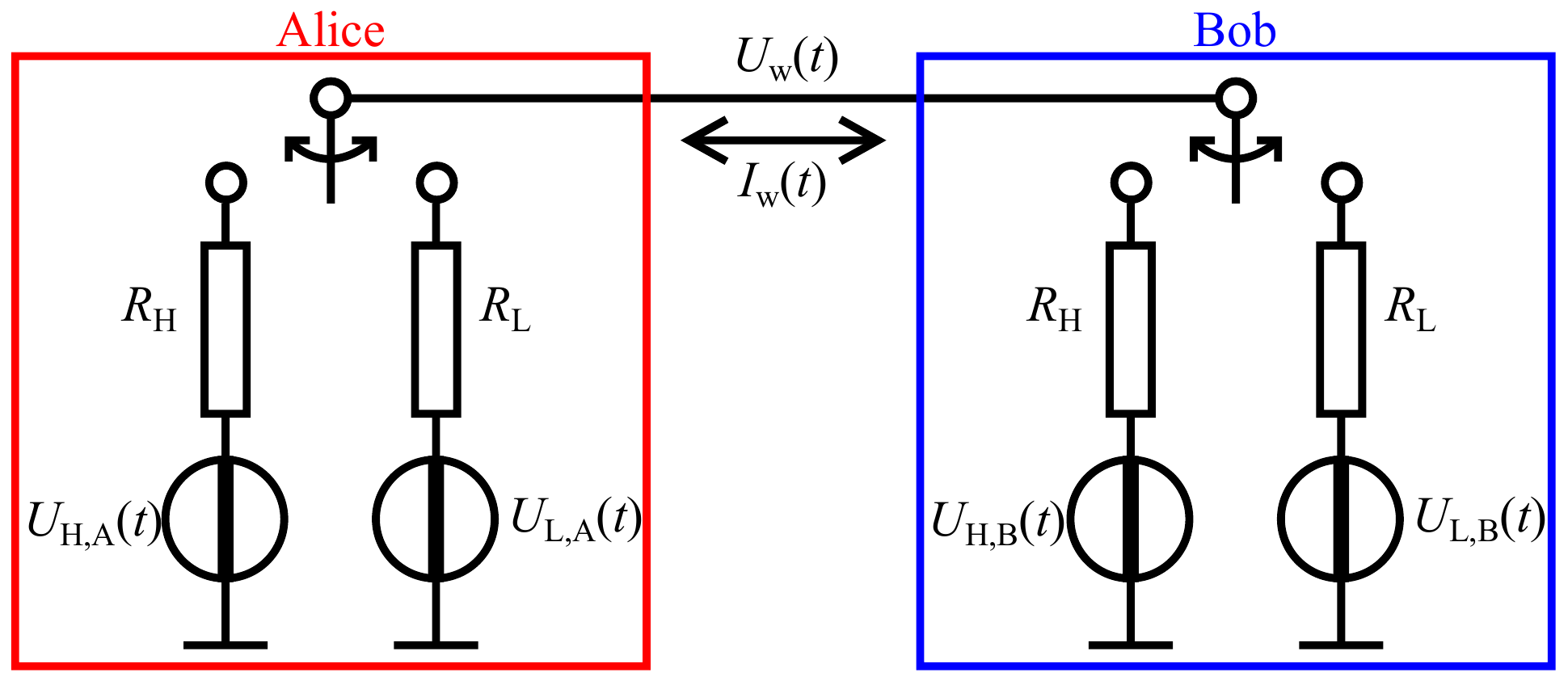}}
\caption{The core of the KLJN scheme. The two communicating parties, Alice and Bob, are connected via a wire. The wire voltage and current are denoted as $U_\mathrm{w}(t)$ and $I_\mathrm{w}(t)$, respectively. The parties have identical pairs of resistors $R_\mathrm{H}$ and $R_\mathrm{L}$ ($R_\mathrm{H} > R_\mathrm{L}$) that are randomly selected and connected to the wire at the beginning of the bit exchange period. The statistically independent thermal noise voltages $U_\mathrm{H,A}(t)$, $U_\mathrm{L,A}(t)$, and $U_\mathrm{H,B}(t)$, $U_\mathrm{L,B}(t)$ represent the noise voltages of the resistors $R_\mathrm{H}$ and $R_\mathrm{L}$ of Alice and Bob, respectively.}
\label{figKLJN}
\end{figure}

Fig.~\ref{figKLJN} illustrates the core of the KLJN scheme. The two communicating parties, Alice and Bob, are connected via a wire. They have identical pairs of resistors, $R_\mathrm{A}$ and $R_\mathrm{B}$. The statistically independent thermal noise voltages $U_\mathrm{H,A}(t)$, $U_\mathrm{L,A}(t)$, and $U_\mathrm{H,B}(t)$, $U_\mathrm{L,B}(t)$ represent the noise voltages of the resistors $R_\mathrm{H}$ and $R_\mathrm{L}$ ($R_\mathrm{H} > R_\mathrm{L}$) of Alice and Bob, respectively, which are generated from random number generators (RNGs).

At the beginning of each bit exchange period (BEP), Alice and Bob randomly choose one of their resistors to connect to the wire. The wire voltage $U_\mathrm{w}(t)$ and current $I_\mathrm{w}(t)$ are as follows:

\begin{equation}
U_\mathrm{w}(t)=I_\mathrm{w}(t)R_\mathrm{B}+U_\mathrm{B}(t),\label{Uw}
\end{equation}

\begin{equation}
I_\mathrm{w}(t)=\frac{U_\mathrm{A}(t)-U_\mathrm{B}(t)}{R_\mathrm{A}+R_\mathrm{B}}\label{Iw}
\end{equation}

\noindent where $U_\mathrm{A}(t)$ and $U_\mathrm{B}(t)$ denote the instantaneous noise voltage of the resistor chosen by Alice and Bob, respectively. Alice and Bob (as well as Eve) use the mean-square voltage of the wire to assess the bit status, given by the Johnson formula

\begin{equation}
U_{\mathrm{w,eff}}^2=4kT_{\mathrm{eff}}R_{\mathrm{p}}\Delta{f_{\mathrm{B}}}\label{meansquare}
\end{equation}

\noindent where $k$ is the Boltzmann constant (1.38 x 10\textsuperscript{-23} J/K), $T_\mathrm{eff}$ is the publicly agreed effective temperature, $R_\mathrm{P}$ is the parallel combination of Alice and Bob’s chosen resistors, given by

\begin{equation}
R_{\mathrm{p}}=\frac{R_{\mathrm{A}}R_\mathrm{B}}{R_\mathrm{A}+R_\mathrm{B}},\label{Rp}
\end{equation}

\noindent and $\Delta$$f_\mathrm{B}$ is the bandwidth of the cable.

\begin{figure}[htbp]
\centerline{\includegraphics[width=8.25cm]{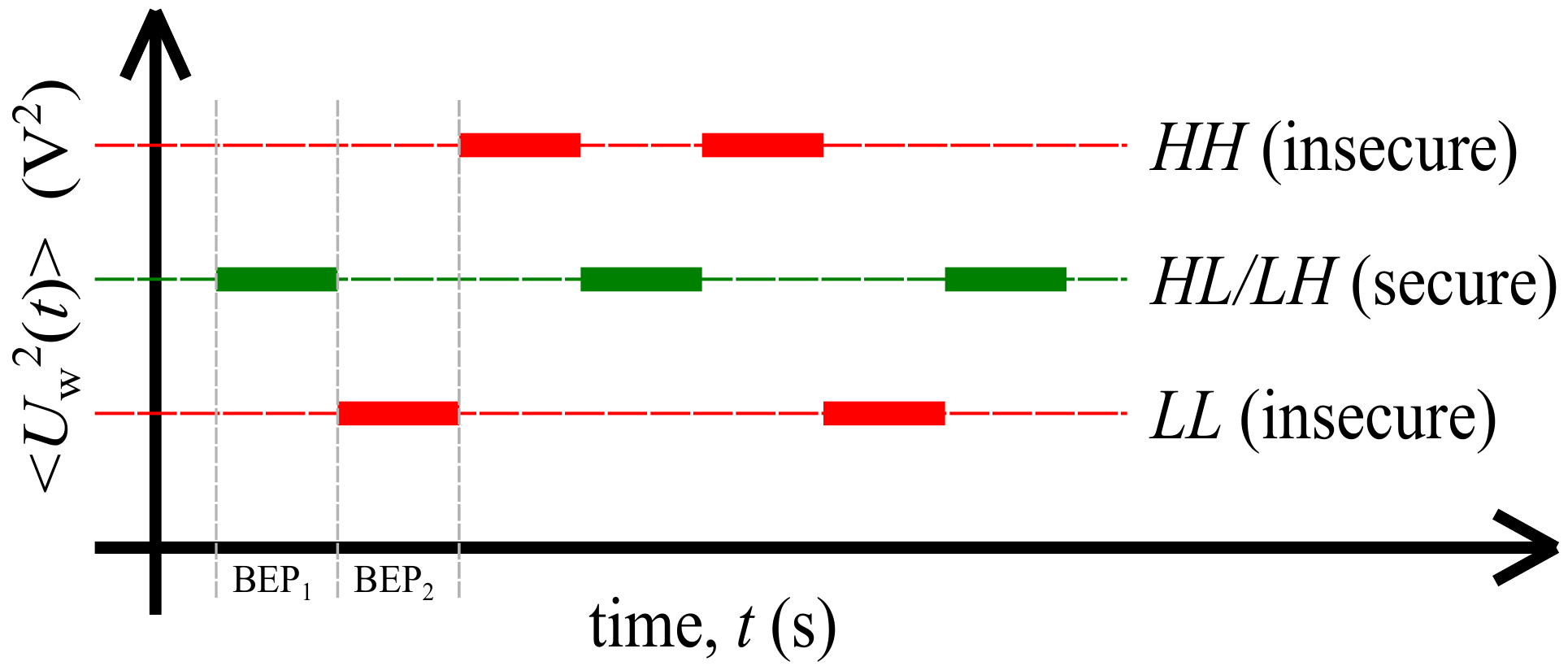}}
\caption{The three mean-square voltage levels. The \textit{HH} and \textit{LL} cases represent insecure situations because they form a unique mean-square voltage. The \textit{HL} and \textit{LH} cases represent secure bit exchange because Eve cannot distinguish between the corresponding two resistance situations while Alice and Bob can because they know their own connected resistance values.}
\label{threelevels}
\end{figure}

Four possible resistance situations can be formed by Alice and Bob: \textit{HH}, \textit{LL}, \textit{LH}, and \textit{HL}. Using the Johnson formula, these correspond to three mean-square voltage levels, as shown in Fig.~\ref{threelevels}. 

The \textit{HH} and \textit{LL} cases represent insecure situations because they form a unique mean-square voltage. These insecure exchange values are discarded by Alice and Bob. The \textit{HL} and \textit{LH} cases represent secure bit exchange because Eve cannot distinguish between the corresponding two resistance situations (\textit{LH} or \textit{HL}) while Alice and Bob can.

Several attacks against the KLJN system have been proposed \cite{b73,b74,b75,b76,b77,b78,b79,b80,b81,b82,b83,b84,b85,b86,b87,b88,b89,b90,b91,b92}, but no attack has been able to compromise its information-theoretic security because each known attack is either invalid or it can be nullified by a corresponding defense scheme. The attack presented in this paper is based on the assumption that the random number generators Alice and Bob use to generate their noises are compromised.

\subsection{Random Number Generator Attacks}
\label{RNG}

There are two classes of practical random number generators: true (physical) and computational. The nature of computational RNGs is that they collect randomness from various low-entropy input streams and try to generate outputs that are in practice indistinguishable from truly random streams \cite{b93,b94,b95,b96,b97,b98}. The randomness of an RNG relies on the uncertainty of the random seed, or initialization vector, and a long sequence with uniform distribution. The moment an adversary learns the seed, the outputs are known, and the RNG is compromised.

Various RNG attacks exist against conditionally secure communications \cite{b93,b94,b95,b96,b97,b98}. Unconditionally secure communications also require true random numbers for perfect security, and that is also true for the noises of Alice and Bob and for the randomness of their switch driving, yet it is unclear how Eve can utilize compromised RNGs to attack the KLJN scheme. Here we demonstrate with simple attack examples that compromised noises lead to information leak. 

The rest of this paper is organized as follows. Section~\ref{experiment} describes two new attack protocols, Section~\ref{results} demonstrates the results, and Section~\ref{conclusion} concludes this paper.

\section{Attack Methodology}
\label{experiment}

Two situations are introduced where Eve can use compromised RNGs to crack the KLJN scheme: one where Eve knows the seed of both Alice’s and Bob’s generators (bilateral parameter knowledge), and another where Eve knows the seed of only Alice’s generator (unilateral parameter knowledge).

\subsection{Bilateral Parameter Knowledge}
\label{bilateral}
Eve knows the seed of both Alice’s and Bob’s RNGs, thus she knows the instantaneous amplitudes of the noise voltage generators for each of their resistors (see Fig.~\ref{figKLJN}). With $U_{\mathrm{H,A}}(t)$, $U_{\mathrm{L,A}}(t)$, $U_{\mathrm{H,B}}(t)$, and $U_{\mathrm{L,B}}(t)$ known, Eve uses \eqref{Uw} and \eqref{Iw} to come up with the four possible waveforms for $U_{\mathrm{w}}(t)$ and $I_{\mathrm{w}}(t)$ and therefore the four possible waveforms for the instantaneous power flow from Alice to Bob, given by

\begin{equation}
P_\mathrm{w}(t)=U_\mathrm{w}(t)I_\mathrm{w}(t).\label{Pw}
\end{equation}

\noindent Eve measures $U_{\mathrm{w}}(t)$ and $I_{\mathrm{w}}(t)$ and determines the actual $P_{\mathrm{w}}(t)$, which will be identical to one of her four hypothetical waveforms.

Note, the protocol shown above can be simplified to the extreme: Eve's instruments may have just one bit of resolution, which is suitable to determine solely the instantaneous direction of the power flow.

\subsection{Unilateral Parameter Knowledge}
\label{unilateral}
Eve knows only the seed of Alice’s generator RNG, thus she knows merely the noise generator outputs of Alice’s resistors, $U_{\mathrm{L,A}}(t)$ and $U_{\mathrm{H,A}}(t)$. Bob’s generator voltages are unknown to her. Eve measures the wire voltage $U_{\mathrm{w}}(t)$ and wire current $I_{\mathrm{w}}(t)$. Then, from $I_\mathrm{w}(t)$, she calculates the hypothetical voltage drops on Alice's possible choice of resistances $R_{\mathrm{H}}$ and $R_{\mathrm{L}}$. With these data, she tests two hypotheses:\\

\begin{description}[leftmargin=!,labelwidth=\widthof{The longest label}]
\item [Hypothesis (i):] Alice has chosen $R_{\textrm{L}}$.
\item [Hypothesis (ii):] Alice has chosen $R_{\textrm{H}}$.\\
\end{description}

With Hypothesis (i), Eve takes the sum

\begin{equation}
U_{\mathrm{R_L}}^\ast(t)=U_{\mathrm{w}}(t)+I_{\mathrm{w}}(t)R_{\mathrm{L}}.\label{UA}
\end{equation}

\noindent To test Hypothesis (i) she compares the $U_{\mathrm{R_L}}^\ast(t)$ determined by \eqref{UA} to $U_{\mathrm{L,A}}(t)$. If they are identical then hypothesis (i) is correct, otherwise hypothesis (ii) is valid.

Finally, Eve evaluates the measured mean-square voltage on the wire over the bit exchange period. From that value, by using \eqref{meansquare}, she evaluates the parallel resultant $R_{\mathrm{P}}$ of the resistances of Alice and Bob. From $R_{\mathrm{P}}$ and $R_{\mathrm{A}}$, she calculates $R_{\mathrm{B}}$ by \eqref{Rp}, thus she has cracked the KLJN scheme.

\section{Demonstration}
\label{results}

\subsection{Johnson Noise Emulation}

First, we generated Gaussian band-limited white noise (GBLWN). Precautions were used to avoid aliasing errors, improve Gaussianity, and reduce bias:

\begin{enumerate}[label=(\roman*)]
\item At first, using the MATLAB randn() function, 2\textsuperscript{24} or 16,777,216 Gaussian random numbers were generated.

\item This process was repeated 10 times to generate 10 independent raw noise series, and then an ensemble average was taken out of those 10 series to create one single noise time function with improved Gaussianity and decreased short-term bias.

\item Then this time series was converted to the frequency domain by a Fast Fourier Transformation (FFT). To get rid of any aliasing error, we opened the real and imaginary portions of the FFT spectrum and doubled their frequency bandwidths by zero padding to represent Nyquist sampling.

\item Finally, we performed an inverse FFT (IFFT) of the zero-padded spectrum to get the time function of the anti-aliased noise.
\end{enumerate}

\noindent The real portion of the IFFT result is the band-limited, anti-aliased noise with improved Gaussianity and decreased bias.

\begin{figure}[htbp]
\centerline{\includegraphics[width=8.25cm]{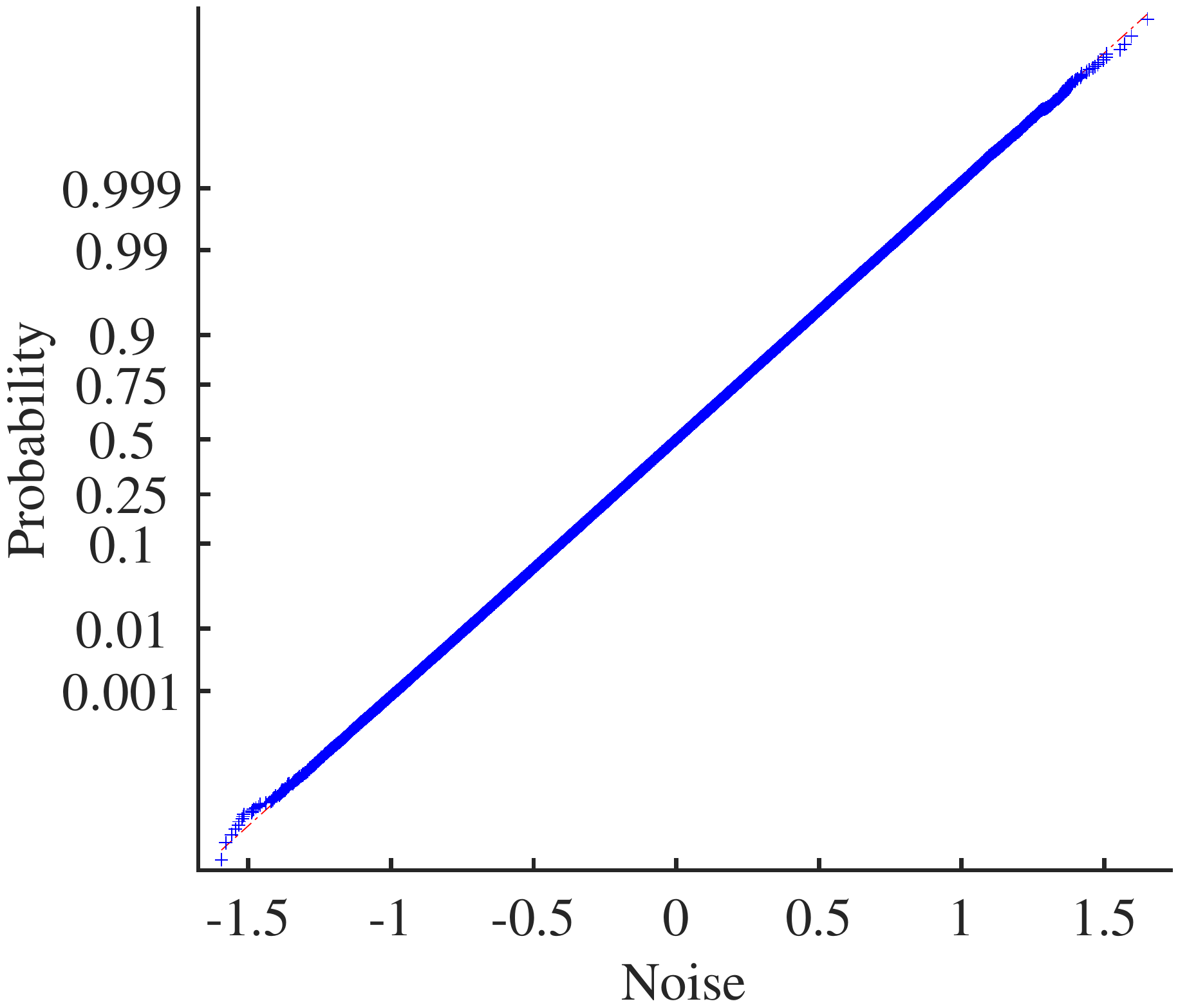}}
\caption{Normal-probability plot of the noise. A straight line indicates a pure Gaussian distribution.}
\label{probplot_noise}
\end{figure}

\begin{figure}[htbp]
\centerline{\includegraphics[width=8.25cm]{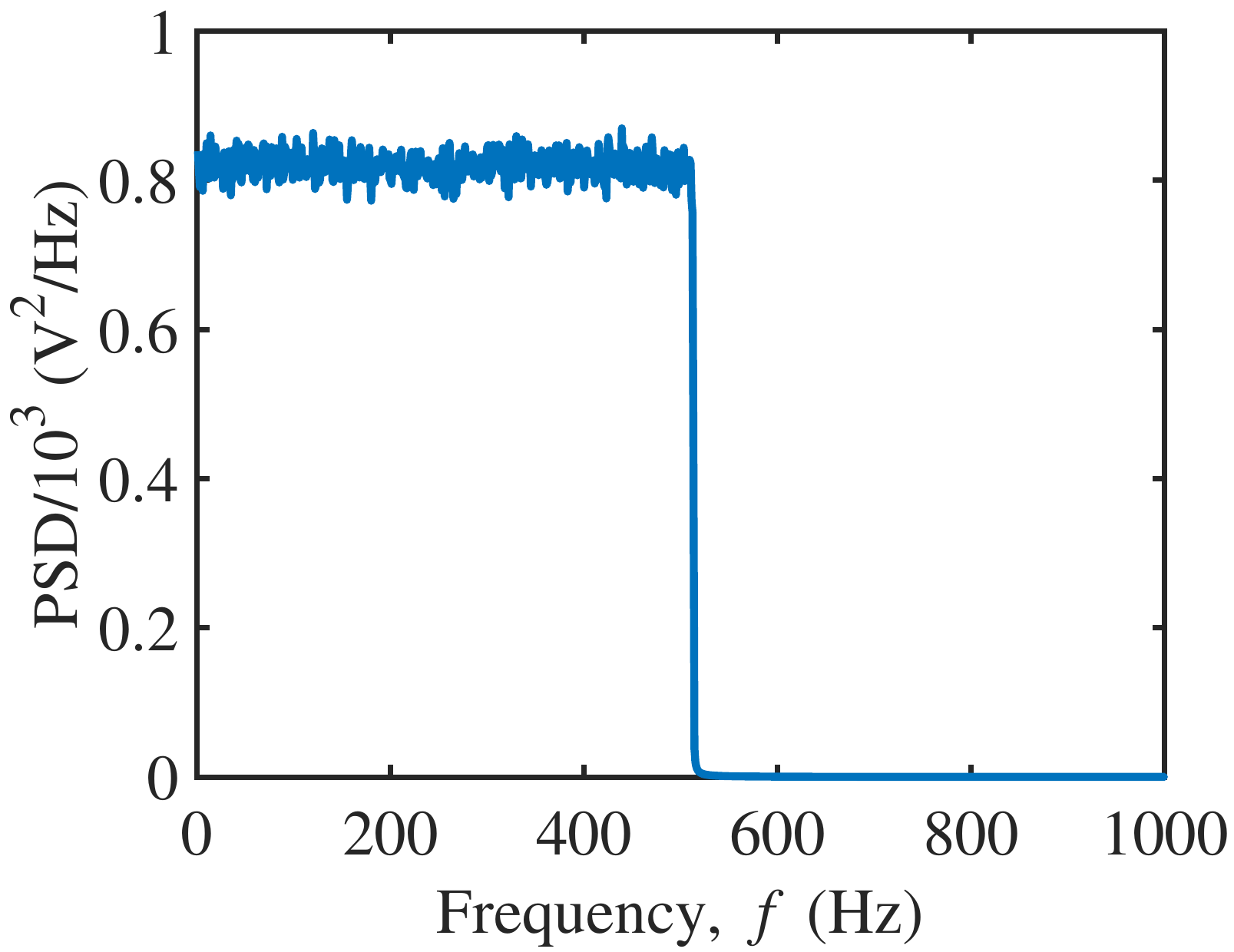}}
\caption{Power spectral density of the noise. The bandwidth of the noise is 500 Hz, see \eqref{meansquare}.}
\label{PSD_noise}
\end{figure}

The probability plot of the generated noise is shown in Fig.~\ref{probplot_noise}, showing that the noise is Gaussian. Fig.~\ref{PSD_noise} demonstrates that the noise has a band-limited, white power density spectrum and that it is anti-aliased.

The final step was to convert this normalized Gaussian noise to Johnson noise of physical values. Two separate independent noise files were created, by using the methods described above. Then their effective (RMS) values were scaled by the Johnson formula \eqref{meansquare} to the values relevant for the actual $T_\mathrm{eff}$, $R_\mathrm{L}$, $R_\mathrm{H}$, and $\Delta$$f_\mathrm{B}$. We chose $T_\mathrm{eff}$=10\textsuperscript{18} K, $R_\mathrm{L}$=10 k$\Omega$, $R_\mathrm{H}$=100 k$\Omega$, and $\Delta$$f_\mathrm{B}$=500 Hz. From the Nyquist Sampling Theorem

\begin{equation}
\tau=\frac{1}{2\Delta f_\mathrm{B}}\label{Nyquist}
\end{equation}

\noindent where $\tau$ represents the time step, a $\Delta$$f_\mathrm{B}$ of 500 Hz renders a time step of 10\textsuperscript{-3} seconds. Because this is ideal band-limited white noise, the Nyquist samples are statistically independent.

\subsection{Attack Demonstration when Eve Knows Both Noises}
\label{bidem}
\begin{figure}[htbp]
\centerline{\includegraphics[width=8.25cm]{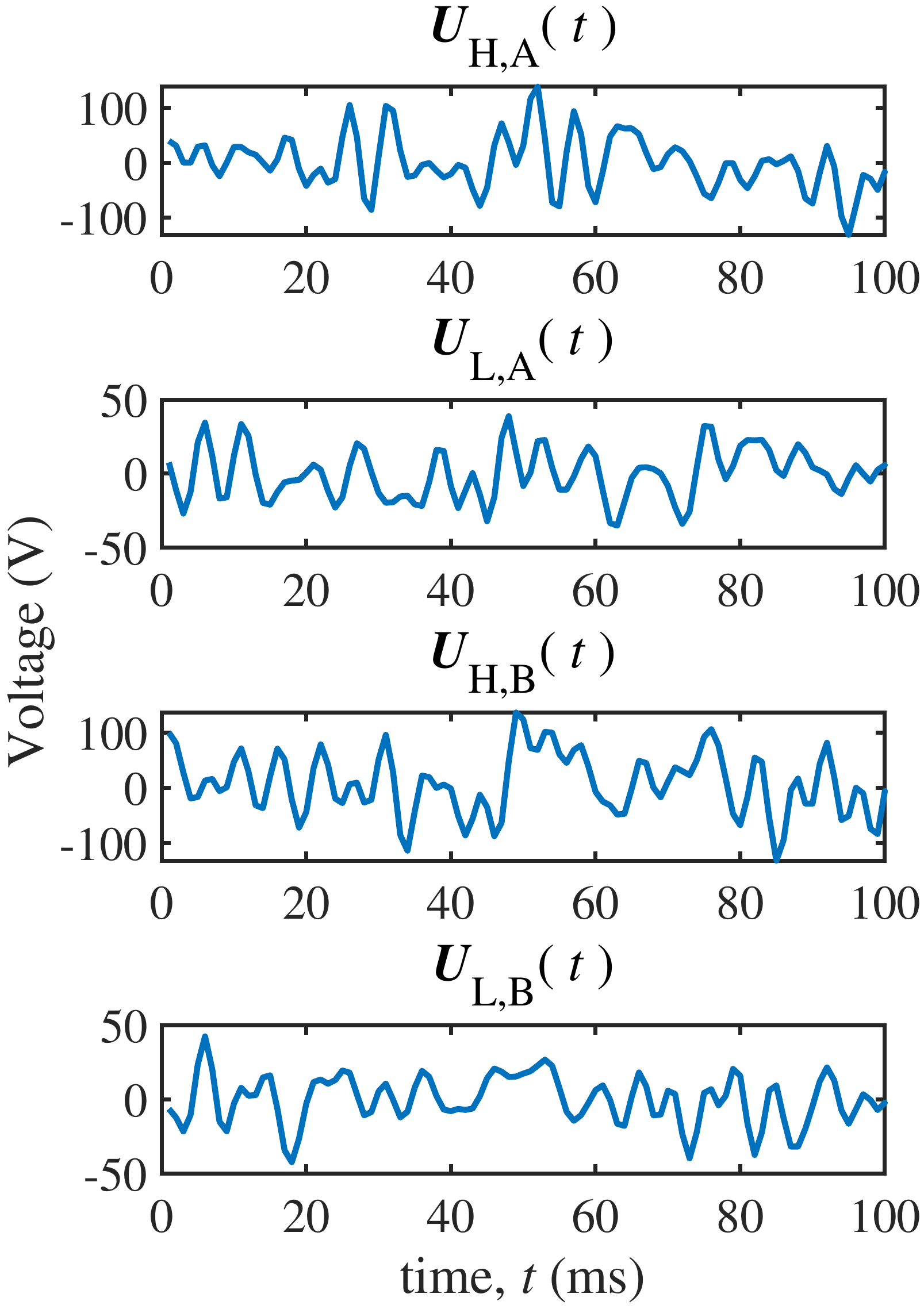}}
\caption{A realization of $U_\mathrm{H,A}(t)$, $U_\mathrm{L,A}(t)$, $U_\mathrm{H,B}(t)$, and $U_\mathrm{L,B}(t)$ (see Fig.~\ref{figKLJN}) displayed over 100 milliseconds. $U_\mathrm{H,A}(t)$ is the noise voltage of Alice’s $R_\mathrm{H}$, $U_\mathrm{L,A}(t)$ is the noise voltage of Alice’s $R_\mathrm{L}$, $U_\mathrm{H,B}(t)$ is the noise voltage of Bob’s $R_\mathrm{H}$, and $U_\mathrm{L,B}(t)$ is the noise voltage of Bob’s $R_\mathrm{L}$. Each time step is one millisecond.}
\label{allvoltages}
\end{figure}

At the bilateral parameter knowledge (see Section~\ref{experiment}-\ref{bilateral}), a realization of the noise amplitudes over 100 milliseconds is displayed in Fig.~\ref{allvoltages}.

\begin{figure}[htbp]
\centerline{\includegraphics[width=8.25cm]{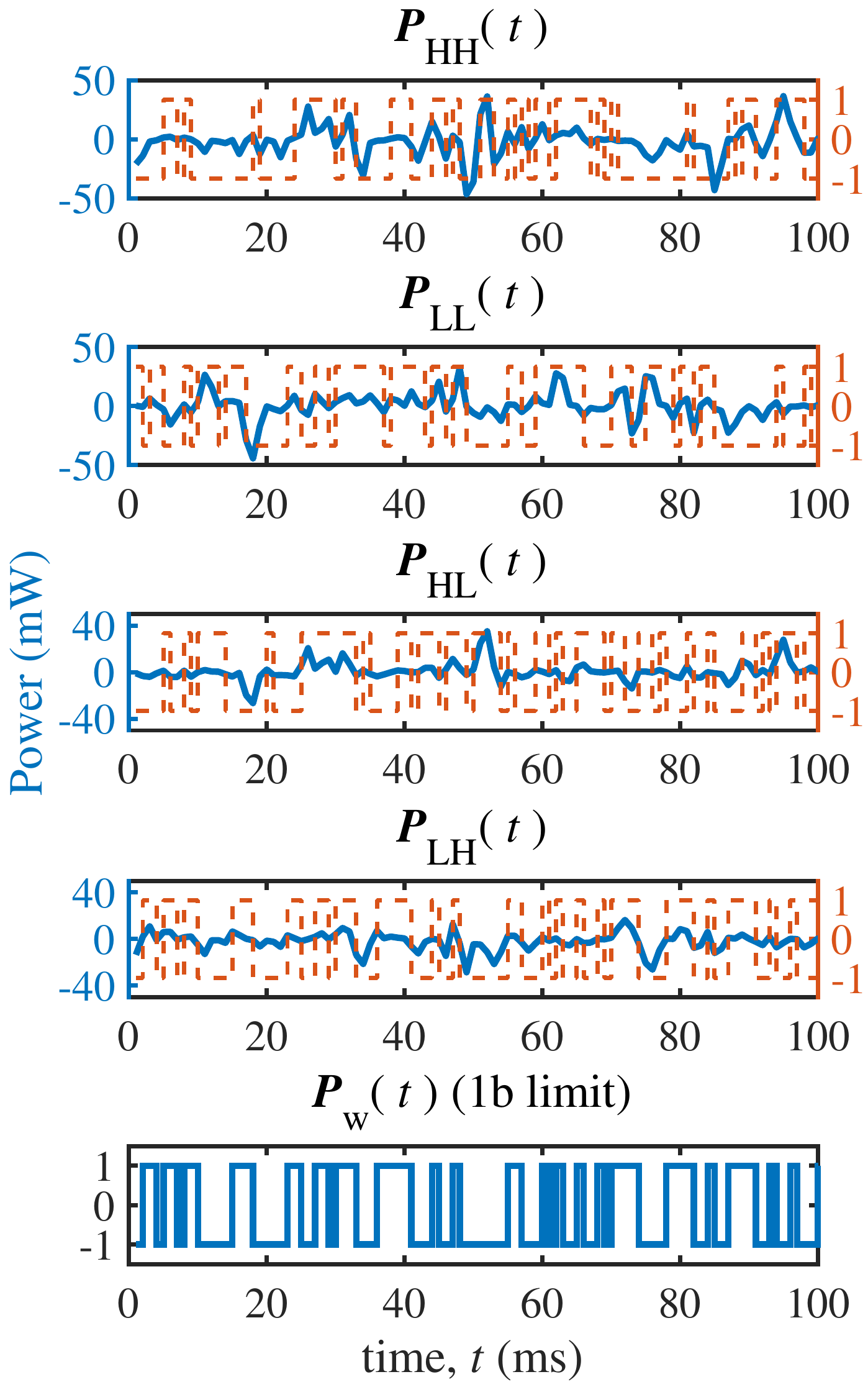}}
\caption{The hypothetical and their 1-bit limit waveforms for $P_\mathrm{w}(t)$ generated by Eve. $P_\mathrm{HH}(t)$ is the hypothetical power flow in the \textit{HH} case, $P_\mathrm{LL}(t)$ is the hypothetical power flow in the \textit{LL} case, $P_\mathrm{HL}(t)$ is the hypothetical power flow in the \textit{HL} case, and $P_\mathrm{LH}(t)$ is the hypothetical power flow in the \textit{LH} case. The 1-bit limit waveforms are represented by the orange dashed lines. $P_\mathrm{w}(t)$ is the single-bit measurement of the actual power flow.}
\label{1beve}
\end{figure}

Fig.~\ref{1beve} shows the hypothetical power flow waveforms generated by Eve (top four graphs). The single-bit plot of her measurement of the actual power flow $P_\mathrm{w}(t)$ is shown by the bottom graph. If the power is delivered from Alice to Bob, the single-bit value is +1, while in the opposite case the result is -1. The orange dashed lines in the upper plots indicate the power flow in the 1-bit resolution limit. At the present situation, Eve’s measured single-bit data are identical to the dashed lines in the \textit{LH} scenario only, thus Eve decides that \textit{LH} is the secure resistance situation.

However, similarly to the case of string verification with noise-based logic \cite{b99}, two independent random bit sequences, such as two different binary sequences in Fig.~\ref{1beve}, may run identically for $n$ subsequent steps with probability

\begin{equation}
P=\frac{1}{2^n}\label{probability}
\end{equation}

\noindent where the independence of the subsequent samples within a given binary sequence is also essential, and $n=t/\tau$.

While such an identical match between two of the 4 noises is taking place, Eve cannot decide which one of the hypothetical sequences is valid thus the actual resistor situation (see Fig.~\ref{threelevels}) remains secure. However, the exponential decay of the probability in \eqref{probability} yields an efficient cracking of the secure key bit value within a short time. In accordance with \eqref{probability}, the probability of two independent ones of our binary sequences (see Fig.~\ref{1beve}) running identically is

\begin{equation}
P(t)\approx\frac{1}{2^{t/\tau}}=\frac{1}{2^{1000t}}\label{poft}
\end{equation}

\noindent The approximation sign is due to the quantized nature of $P(t)$ because it is staying constant during the bit exchange periods.

\begin{figure}[htbp]
\centering
\subfloat[]{
	\label{log_securecases}
	\includegraphics[width=4.0cm]{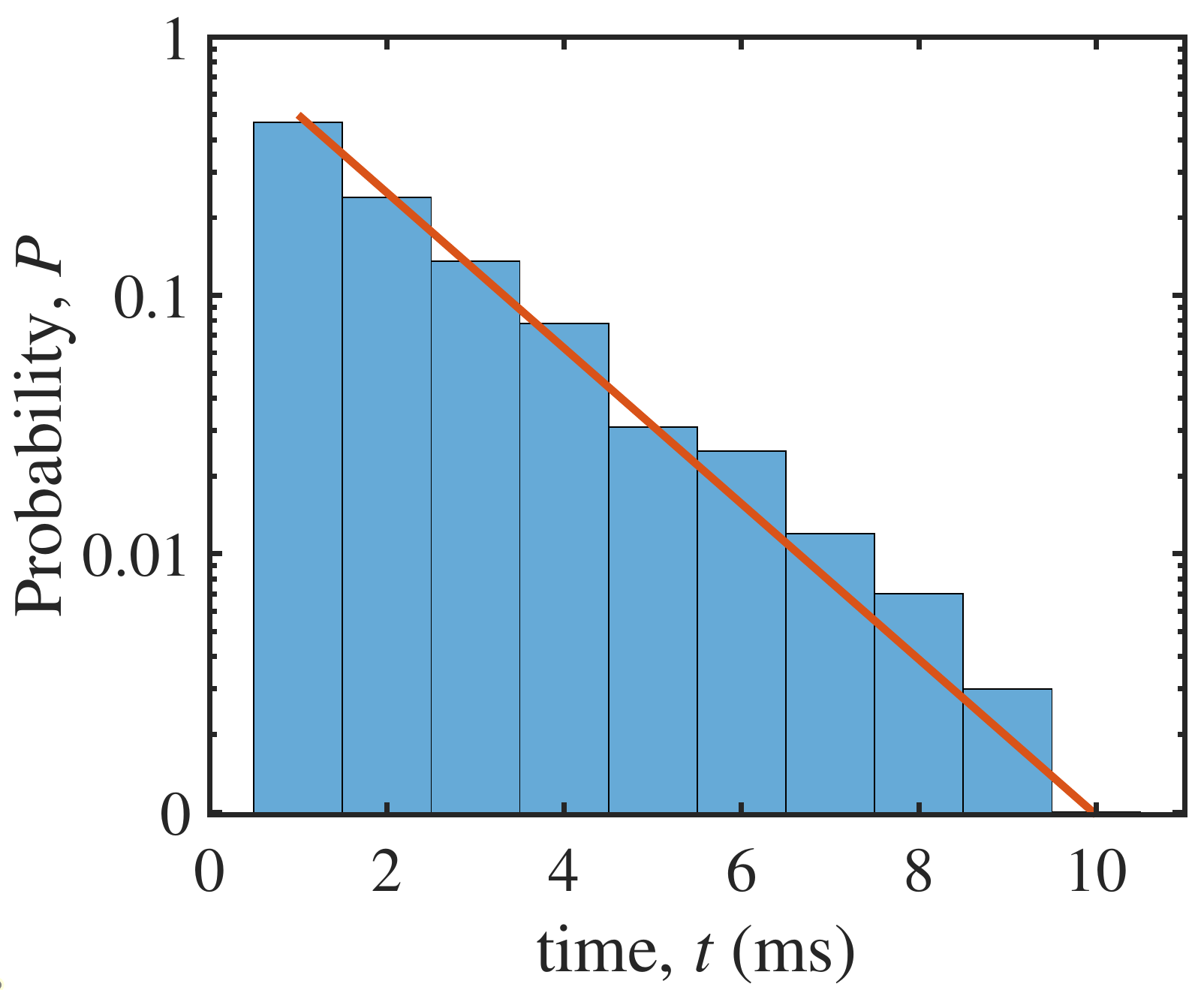} } 
\subfloat[]{
	\label{log_crack}
	\includegraphics[width=4.1cm]{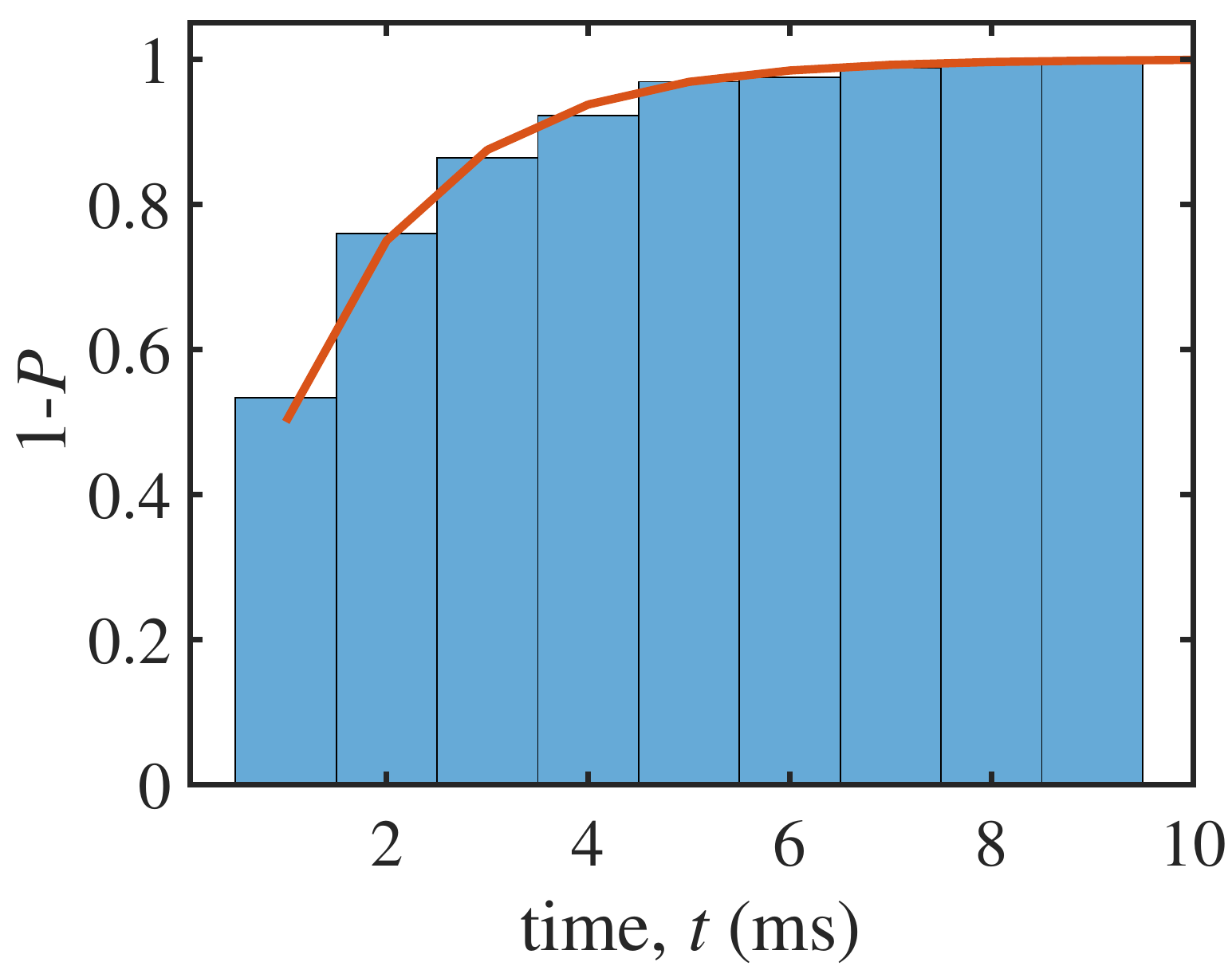} } 
\caption{\ Probabilities during the attack with single-bit resolution of Eve's measurements. (a) Probability that the exchange of the current key bit is still secure, and (b) probability that Eve has already cracked it. The histograms represent the simulation results, and the orange lines represent the scaling given by \eqref{poft}.}
\end{figure}

This simulation was run 1000 times. Fig.~\ref{log_securecases} shows the probability that the exchange of the current key bit is still secure. Fig.~\ref{log_crack} shows the probability that Eve has already cracked the bit. The orange lines follow the scaling given by \ref{poft}.

\subsection{Attack Demonstration when Eve Knows Only One of the Sources}
\label{unidem}
\begin{figure}[htbp]
\centerline{\includegraphics[width=8.25cm]{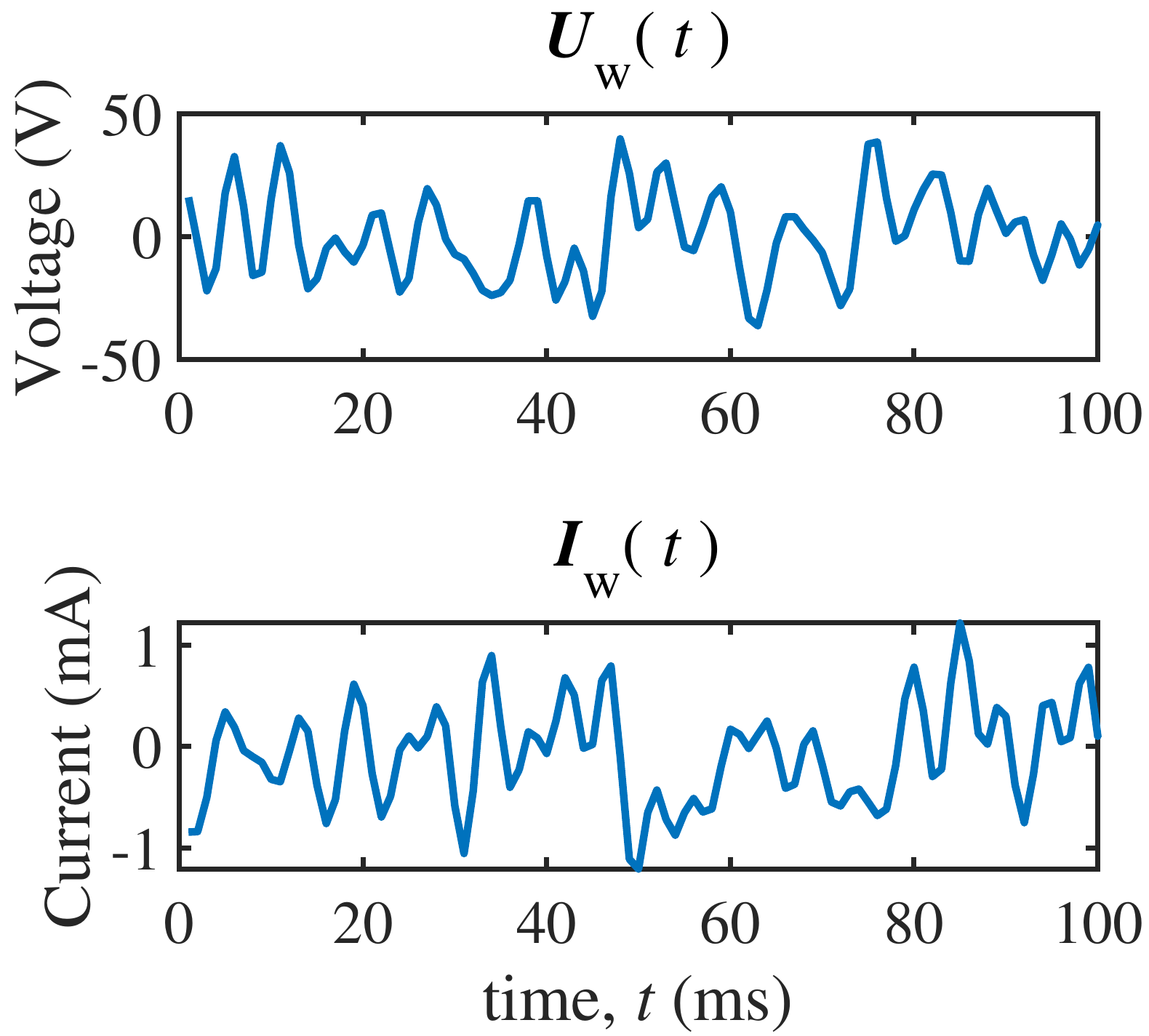}}
\caption{A realization of $U_\mathrm{w}(t)$ and $I_\mathrm{w}(t)$ (see Fig.~\ref{figKLJN}) for the \textit{LH} situation is displayed over 100 milliseconds. Eve measures and records these data.}
\label{Eve_wire_data}
\end{figure}

\begin{figure}[htbp]
\centerline{\includegraphics[width=8.25cm]{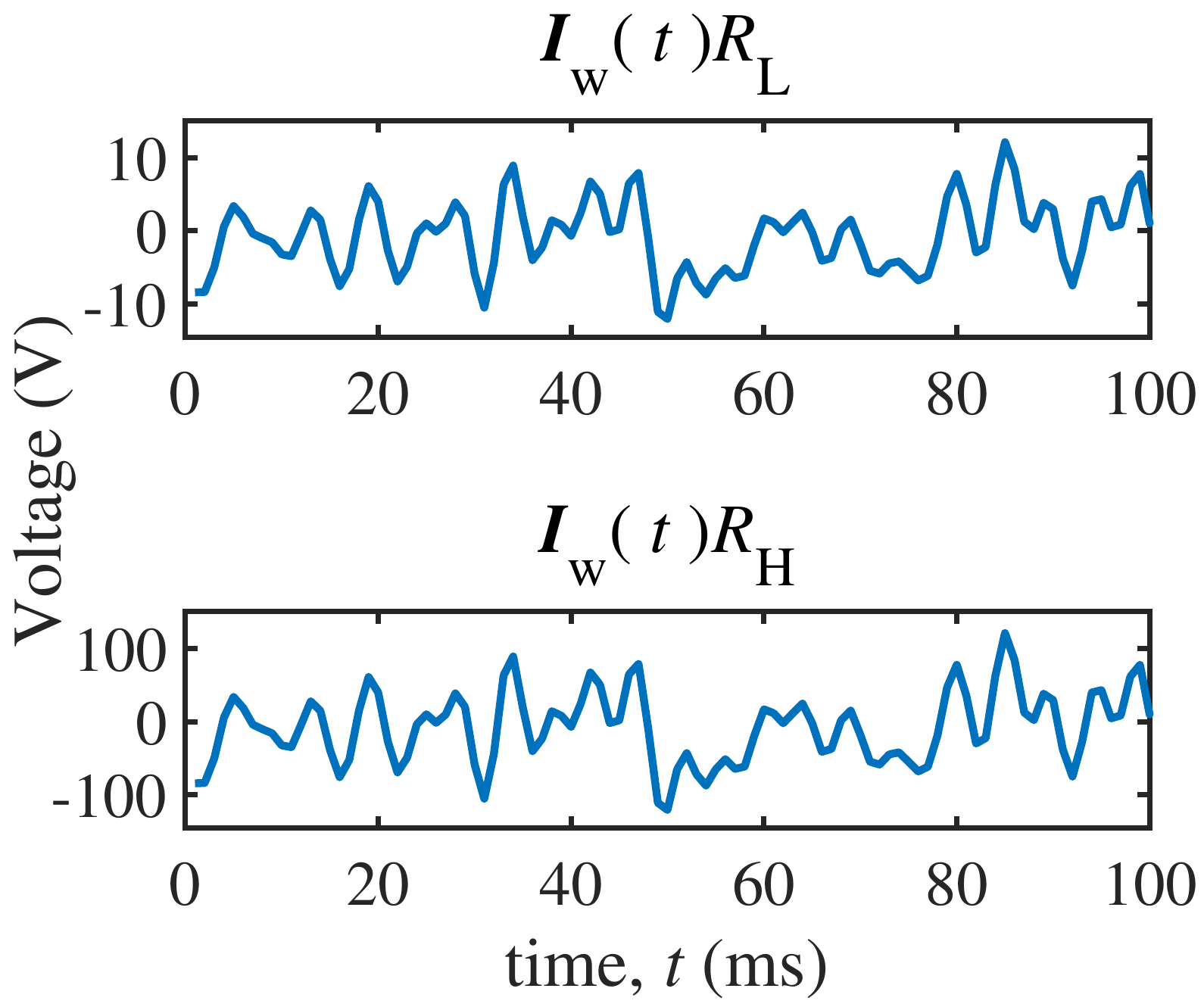}}
\caption{Hypothetical noise voltage drops across $R_\mathrm{L}$ and $R_\mathrm{H}$ by Eve's current measurements and Ohm's law. These results are used in \eqref{UA} to calculate the hypothetical waveforms for $U_\mathrm{L,A}(t)$ and $U_\mathrm{H,A}(t)$.}
\label{Possible_voltagedrops}
\end{figure}

At the unilateral parameter knowledge (see Section~\ref{experiment}-\ref{unilateral}), Eve knows the seed of Alice’s RNG, while the seed of Bob’s RNG is unknown to her, thus we suppose she knows $U_\mathrm{A}(t)$ (like at the bilateral case) but not $U_\mathrm{B}(t)$. A realization of the wire voltage and current, $U_\mathrm{w}(t)$ and $I_\mathrm{w}(t)$, under the \textit{LH} condition over 100 milliseconds is displayed in Fig.~\ref{Eve_wire_data}. Fig.~\ref{Possible_voltagedrops} shows the hypothetical noise voltages generated by Eve's simulations across $R_\mathrm{L}$ and $R_\mathrm{H}$ (see \eqref{UA}).

\begin{figure}[htbp]
\centerline{\includegraphics[width=8.25cm]{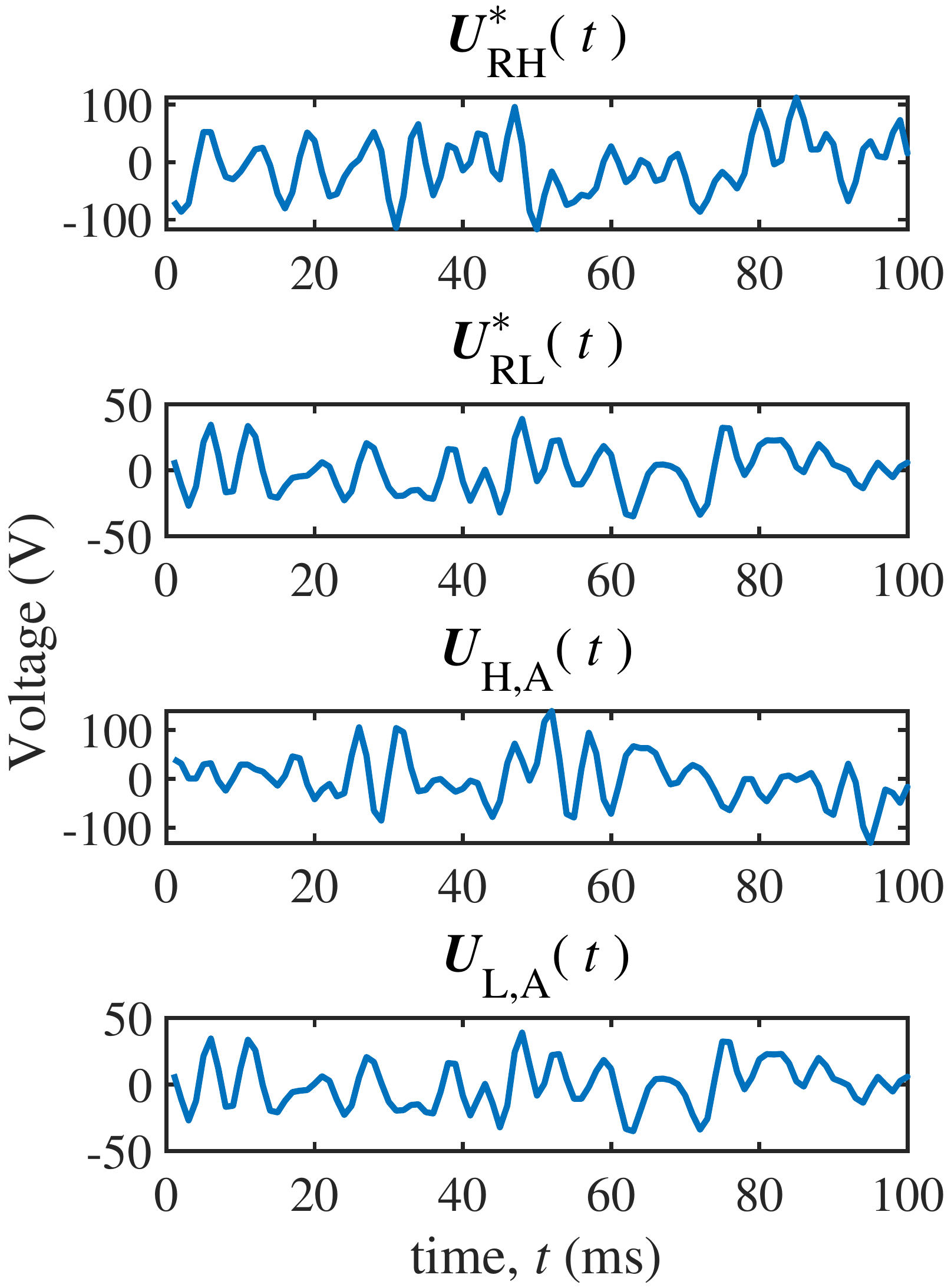}}
\caption{Eve has calculated the hypothetical $U_\mathrm{R_H}^\ast(t)$ and $U_\mathrm{R_L}^\ast(t)$ (see \eqref{UA}) and she has her waveforms for $U_\mathrm{H,A}(t)$ and $U_\mathrm{L,A}(t)$. $U_\mathrm{R_L}^\ast(t)$ matches $U_\mathrm{L,A}(t)$ thus Alice has connected $R_\mathrm{L}$.}
\label{Evecalc_vs_UA}
\end{figure}

Fig.~\ref{Evecalc_vs_UA} shows Eve’s results for the hypothetical $U_\mathrm{R_L}^\ast(t)$ and $U_\mathrm{R_H}^\ast(t)$ (see \eqref{UA}) in comparison with her waveforms for $U_\mathrm{L,A}(t)$ and $U_\mathrm{H,A}(t)$. In this case, the waveform for $U_\mathrm{R_L}^\ast(t)$ is identical to that of $U_\mathrm{L,A}(t)$, thus Eve decides that Alice has chosen $R_\mathrm{L}$.

Finally, Eve evaluates the measured mean-square voltage on the wire over the bit exchange period. From that value, by using \eqref{meansquare}, she evaluates the parallel resultant $R_\mathrm{P}$ of the resistances of Alice and Bob. From $R_\mathrm{P}$ and $R_\mathrm{A}$, she can calculate $R_\mathrm{B}$. In this particular case, from the mean-square voltage Eve will learn that the actual situation is \textit{LH} thus Bob has chosen $R_\mathrm{H}$ because Alice has $R_\mathrm{L}$.

\section{Conclusion}
\label{conclusion}
Secure key exchange protocols utilize random numbers, and compromised random numbers lead to information leak. So far, it had been unknown how Eve can utilize compromised random number
generators to attack the KLJN protocol. To demonstrate how compromised RNGs can be utilized by Eve, we have introduced two simple attacks on the KLJN scheme.

We showed that if Eve knows the seed of both Alice's and Bob's RNGs, that is, when she exactly knows the random numbers, she can crack the bit exchange even if her measurements have only one bit of resolution. The situation is similar to string verification in the noise-based logic systems. The cracking of the exchanged bit is exponentially fast; Eve can extract the bit within a fraction of the bit exchange period, thus Eve will learn the exchanged bit faster than Alice and Bob who know only their own random numbers.

We have also shown that if Eve knows the seed of only Alice’s RNG, she can still crack the secure bit, however she is required to utilize the whole bit exchange period.

It is important to note that:
\begin{itemize}
\item This demonstration was done assuming an ideal KLJN scheme. Future work would involve a practical
implementation with a cable simulator and related delays and transients.
\item A deterministic knowledge of the random number(s) by Eve is a strong security vulnerability, however
it is an illustrative way how such attacks can be developed;
\item Open problems are new attack schemes where Eve's knowledge of the RNGs is only statistical.
\end{itemize}

\end{document}